\documentstyle[12pt]{article}

\title{ \vspace*{2.5cm} \bf \Large
	Pseudoscalar Glueball, $\eta'$-meson and its Excitation \\ 
	in the Chiral Effective Lagrangian \vspace*{5 mm}}
	
\author{ M. L. Nekrasov \thanks{ E-mail: {\tt nekrasov@mx.ihep.su}}
\smallskip  \\
{\small Institute for High Energy Physics, 142284 Protvino, Russia }}

\date{}

\begin{document}

\maketitle

\begin{abstract}

A generalization of the chiral effective lagrangian of order $p^2$ is
proposed which involves the $\eta'$-meson, its excitation, and the
pseudoscalar (PS) glueball. Model-independent constraints are found
for the contributions to the lagrangian of the above singlet states.
Those allow one to independently identify the nature of these singlet
states in the framework of the approach. The mixing among the
iso-singlet states (including $\eta^8$-state) is analysed, and the
hierarchy of the mixing angles is described which is defined by the
chiral and large-$N_c$ expansions. The recent PCAC results are
reproduced, which are related to the problem of the
renormalization-group invariant description of the $\eta'$ and the PS
glueball, and a further analysis of this problem is performed.

\medskip


\end{abstract}

\newpage 

\section{ Introduction}

The problem of the description of glueballs in QCD is longstanding
but still unsolved. In spite of the consensus which has been recently
achieved in the gluonic-lattice calculations \cite{Bali}, there
remains a serious problem of taking into account quark contributions.
The problem seems to be most intricate in pseudoscalar (PS) channel,
where there is the $\eta'$-meson which, being quarkic in its origin,
contains a gluonic contribution that significantly affects its
observable properties \cite{W1,t'H}. Consequently, one may expect
that quarks are quite significant in the formation of the PS
glueball.

In fact, the situation is more complicated than it is usually
assumed, since in the exact QCD the quark-gluon mixing may depend on
the ultraviolet renormalization scale \cite{E-T}. As a rule, this
phenomenon is ignored. However, in some cases that leads to
disastrous effects. Indeed, it has been found in \cite{S-V,K-P} that
the straightforward generalization of the PCAC formula for
$\pi\rightarrow\gamma\gamma$ to $\eta'\rightarrow\gamma\gamma$ is
inconsistent with the renormalization group (RG) and therefore
incorrect in principle. To obtain a correct formula one has to
consider a set of composite operators which mix under RG \cite{S-V}.
A similar investigation was performed which involved both the
$\eta'$-meson and the PS glueball \cite{N1}, and there the RG
invariant composite operators (interpolating fields) were obtained
that generate separately each state. However, that investigation was
ill-fitted for the description of the hadronic decays of the $\eta'$
and the PS glueball. Therefore, the problem needs to be re-analysed
in a more sophisticated fashion.

A systematic approach is based on the chiral effective lagrangian
\cite{Weinberg}. When it is considered in the framework of the chiral
perturbation theory \cite{G-L-1,G-L-2} it allows one to describe
consistently the low-energy interactions of the lightest pseudoscalar
states $\pi,K,\eta$, and their interactions with heavier states
\cite{E-G-P-DR}. The mixing phenomenon, as well, should be tractable
in the framework of this approach. Concerning the mixing between the
PS glueball and pseudoscalar quarkic states, the latter states might
be, first of all, the $\eta'$ and its excitations (including the
radial ones and hybrids). How many states are needed depends on how
heavy the PS glueball is. If its mass lies in the $E/\iota$ range
(1.4--1.5 GeV) then, most probably, the $\eta'$ and its first radial
excitation would be enough to describe the mixing. If the PS glueball
is heavier then apparently more states would be needed.

Of course, with singlet states, the chiral symmetry is no longer
sufficient to constrain their contributions to the chiral effective
lagrangian (except the case of $\eta'$ which is conditioned by its
U$(1)_{A}$ transformation property \cite{W1,G-L-2}). Moreover, the
chiral symmetry does not allow one to distinguish between different
singlet states. Nevertheless, one may expect that if the dynamical
nature of the singlet states is different then some extra dynamical
(not symmetry) conditions would constrain the singlet-state
contributions to the lagrangian. For instance, there is large-$N_c$
expansion which leads to the condition that the gluonic and quarkic
contributions must behave differently at large $N_c$ \cite{W1,W2}.
However, this condition cannot completely suppress any parameter of
the chiral effective lagrangian. Therefore one needs to find some
stronger conditions. The main task of the present paper is to try to
find them. Another task is to reexamine the problem of the RG
invariant description of the singlet states in the framework of the
chiral effective lagrangian.

The structure of the paper is as follows. The next section collects
the necessary notation and discusses the conditions which follow from
the chiral symmetry and RG symmetry in QCD. Section 3 discusses a
generalization of the chiral effective lagrangian which involves
singlet interpolating fields and satisfies the above symmetry
conditions. In Section 4 we search for extra conditions which would
constrain the singlet-state contributions. A consistency condition is
found which additionally constrains the contributions of the $\eta'$
and constrains as well the contributions of the other singlet PS
states. A general necessary condition is proposed for glueballs.
Moreover, a general physical criterion is proposed which allows one
to distinguish between the ground and excited states in the framework
of the approach. Combined together these results allow one to
constrain the contributions of the ground-state PS glueball and an
excitation state over $\eta'$. The structure of the iso-singlet
states mixing is investigated in section 5. Section 6 discusses the
problem of radiative decays of $\eta$ and $\eta'$. Section 7
summarizes the results.

\section{ Symmetry conditions }

A consistent way to introduce chiral effective lagrangian is through
the generating functional \cite{G-L-1,G-L-2}. This method permits to
establish a complete relation between the effective theory and the
underlying theory (QCD). In the path-integral approach the generating
functional may be written in the following equivalent
representations:
\begin{eqnarray}
e^{iW(V,A,S,P,\Theta)} & = & 
\int {\cal D}\left[q,\bar q, G \right] \>
e^{i\int d^4 x {\cal L}_{QCD}(q,\bar q, G; V,A,S,P,\Theta)}       \\
& = & \int {\cal D}\left[U\dots\right]\, e^{i\int
d^4 x {\cal L}_{eff}(U\dots;V,A,S,P,\Theta)}.              \nonumber 
\end{eqnarray} 
Here the first equality defines the generating functional in terms of
QCD parameters with $q$, $\bar q$, and $G$ being the fundamental
fields of quarks and gluons. The quantities $V,A,S,P,\Theta$ are the
sources of the composite operators which generate the states to be
described.  In our case those are the axial-vector and PS quark
currents and their chiral partners.  Besides, we have introduced the
source for the gluon anomaly operator which generates the PS
glueball. The notation is as follows:
\begin{eqnarray}
& &{\cal L}_{QCD} = 
 {\cal L}^0_{QCD} + \bar q\,\gamma_{\mu}\! 
  \left(V_{\mu} \! + \! \gamma_{5} A_{\mu}\right) q -
  \bar q\left(S \! + \! i\gamma_{5}P\right) q + \Theta \mbox{\sf Q},
  \qquad\qquad                                   \\[0.5\baselineskip]
& &V = \sum_{a=0,1,\dots8}\left(\lambda^a/2\right)\,V^a,\quad \dots,
\quad  \mbox{\sf Q} = 
\sqrt{2N_f}\,\frac{\alpha _{s}}{8\pi }\,\frac{1}{2}\,
\epsilon _{\mu \nu \lambda \rho }G^{A\mu \nu }G^{A\lambda \rho} .
                                                            \nonumber
\end{eqnarray}
Here $\lambda^a$ are the flavor Gell-Mann matrices ($\lambda^0 =
\sqrt{2/N_f}\,\mbox{\bf I}$, $N_f=3$). With switched-off sources, $S
= {\rm diag}(m_u,m_d,m_s)$, $P = V = A = \Theta = 0$. The sources may
also be regarded as the external fields which are provided with some
certain transformation properties.

The second equality in (1) presents the generating functional of the
effective theory which copies QCD in terms of the interpolating
fields for observable states. Usually, all heavy states are
considered to be integrated out. Then, the dynamical variables become
only the interpolating fields for the octet of lightest PS states
$\pi,K,\eta$. Since these states may be interpreted as Goldstone
bosons that arise due to dynamical breaking of the chiral symmetry
SU$(3)_{L}\times$SU$(3)_{R}$, their interpolating fields may be
collected in a special unitary $3\times3$ matrix $U$ which under
SU$(3)_{L}\times$SU$(3)_{R}$ transforms as
\begin{equation}
U \> \> \rightarrow \> \Omega_L \, U \, \Omega_R^{\dagger}.
\end{equation}
In the exponential parameterization,
\begin{equation}
U  \> =  \> 
\exp\left(i\sum_{\alpha = 
1,\dots8}\lambda^{\alpha}\eta^{\alpha}\,/\,F\right)
\end{equation}
with $\eta^{\alpha}$ being the interpolating fields and $F$ being a
dimensional parameter (when $\eta^{\alpha}$ are normalized
canonically $F$ is the universal octet decay constant). Under the
`flavor-singlet' chiral group U$(1)_{L}\times$U$(1)_{R}$ matrix $U$
is invariant.

Remember, in QCD the full chiral group U$(3)_{L}\times$U$(3)_{R}$
acts on quarks and on the $\theta$-vacuum. The latter property is
developed in the appearance of the non-vanishing value of $\Theta$
when the U$(1)_A$ chiral rotation of quarks has been performed and
the sources have been switched off:
\begin{equation}
\Theta_{\big\vert\mbox{\scriptsize switched-off}} = \> \omega^0_5 \>.
\end{equation}
Here $\omega^0_5 = (\omega^0_R - \omega^0_L)/2$ is the parameter of
U$(1)_A$ rotation. In spite of U$(3)_{L}\times$U$(3)_{R}$
transformation, lagrangian (2) may be made completely chiral
invariant if one assumes the external fields to transform
simultaneously by
\begin{eqnarray}
L_{\mu} \> \rightarrow  & \Omega_L\,L_{\mu}\,\Omega_L^{\dagger} +
	  i\,\Omega_L\partial_{\mu}\Omega_L^{\dagger}, \qquad &
	  L_{\mu} = V_{\mu} - A_{\mu}                  \nonumber \\
R_{\mu} \> \rightarrow  &\,\Omega_R\,R_{\mu}\Omega_R^{\dagger} +
	  i\,\Omega_R\partial_{\mu}\Omega_R^{\dagger}, \qquad &
	  R_{\mu} = V_{\mu} + A_{\mu}                            \\
M \> \>       \rightarrow   &  \Omega_L\,M\>\Omega_R^{\dagger}, 
				  \,\>\quad\qquad\qquad\qquad &  
		  M \, = S + iP                        \nonumber \\
\Theta \> \>\,\rightarrow   &  \Theta - \omega^0_5 \>\>.
				  \>\>\quad\qquad\qquad\qquad & 
\nonumber  
\end{eqnarray}
Under this condition the transformation property of the generating
functional (1) is only governed by the external chiral anomaly
\cite{Bardin}. The requirement to reproduce this property in the
effective theory leads to the condition that the effective lagrangian
must be a sum of an invariant part and the Wess-Zumino-Witten term
which is responsible for the external anomaly in the effective theory.

Usually, no other QCD symmetry is imposed while constructing ${\cal
L}_{eff}$, because it is assumed that no other symmetry is able to
constrain the effective theory. Nevertheless, as we pointed out in
Introduction, the RG symmetry in QCD may lead to nontrivial
consequences in the singlet channel. Therefore, this symmetry must be
taken into consideration, as well. Earlier, that was not done since
it was not well-known how to renormalize the underlying theory when
the sources for composite operators are switched on. The problem was
solved in \cite{Shore} where it was stated that the renormalized
action of the theory must be extended to include all possible terms
formed from a single composite operator and an arbitrary number of
sources and divergences. Furthermore, Ref. \cite{Shore} showed that
the unit operator $\mbox{\bf 1}$ must be included into the basis of
composite operators, since the renormalized action must include terms
involving sources only. The generating functional, then, becomes RG
invariant if it is understood in terms of the renormalized sources
\cite{Shore}. In our case the latter ones are
\begin{eqnarray}
& \!\!\!\!\!\!\!\!\! & 
S^{a}_{\,R} \> = \> Z_m^{-1} \> S^{a}, \>\quad 
    P^{a}_{\,R} \> = \> Z_m^{-1} \>P^{a},
    \>\quad (a = 0,1,\dots 8)                        \nonumber\\
& \!\!\!\!\!\!\!\!\! & 
V^{\alpha}_{\mu\,R} \> = V^{\alpha}_{\mu}, \quad\quad\>\>\>
    A^{\alpha}_{\mu\,R} = \> A^{\alpha}_{\mu}, 
    \quad\qquad (\alpha = 1,\dots 8)                    \\
& \!\!\!\!\!\!\!\!\! & 
V^{0}_{\mu\,R} \> = V^{0}_{\mu}, \qquad\>\,\,\>
A^0_{\mu\,R} = 
Z^{-1}\,A^0_{\mu} + (1-Z^{-1})\>\partial_{\mu} \Theta, \qquad\>
\Theta_R = \Theta.                                    \nonumber
\end{eqnarray}
Here $Z_m$ and $Z$ are the renormalization constants, index {\small
$R$} shows the renormalized sources (external fields). Note, the
inhomogeneous character of the renormalization of $A^0_{\mu}$ means
that there is renormalization-scale-dependent mixing between the
gluon anomaly operator $\mbox{\sf Q}$ and divergence of the
axial-vector singlet quark current $\mbox{\sf J}^{0}_{\mu 5} = \bar q
\gamma_{\mu } \gamma _{5} (\lambda ^{0}/2)q$,
\begin{equation}
[\mbox{\sf Q}]_R = 
\mbox{\sf Q} - (1 - Z)\>\partial^{\mu}\mbox{\sf J}^0_{\mu 5}, \qquad
[\mbox{\sf J}^0_{\mu 5}]_{R} = Z \, \mbox{\sf J}^{0}_{\mu 5}.
\end{equation}
Besides (7), the auxiliary source of the unit operator must be
renormalized, too \cite{Shore}. However, since the corresponding
formula involves only the terms of the {\it chiral} dimension 4 or
higher, this renormalization rule (and the unit operator itself) may
be disregarded when the effective theory is considered at order $p^2$
of the chiral expansion. So, the sole RG requirement which should be
taken into consideration at order $p^2$ is the requirement that
${\cal L}_{eff}$ must be invariant provided that the external fields
are renormalized by (7).

\section{ Singlet interpolating fields in the chiral effective
          lagrangian }

In what follows we will consider the effective theory in the
framework of the chiral perturbation theory. That allows one to
represent the effective lagrangian in the form of an expansion in
derivatives of fields and quark masses. In case when only octet of
the lightest PS states is involved the explicit form of the
lagrangian is well-known \cite{G-L-2}. At the leading order $p^2$ of
the chiral expansion it is
\begin{eqnarray}
& \!\!\!\!\!\!\!\!\!\!\!\!\!\!\!\!\!\!\!\!\!\!\!\!\!\!
{\cal L}_{eff}\>\,=\>{1 \over 4} F^2
\langle\nabla_{\mu}U\nabla^{\mu}U^{\dagger}\rangle + 
                     {1 \over 2} B F^2
\langle M_{\Theta}U^{\dagger}+M_{\Theta}^{\dagger}U\rangle +
{1 \over 2} H\,\nabla_{\!\mu}\Theta \nabla^{\mu}\Theta &     \\[1mm]
& \!\! \nabla_{\mu} U \> = \>
\partial_{\mu} U - i\widetilde{L}_{\mu}U +iU\widetilde{R}_{\mu},  
                                                         \>\quad
\nabla_{\!\mu}\Theta =  \partial_{\mu}\Theta - A^0_{\mu},\>\quad
M_{\Theta} = (S+iP)\,e^{i\lambda^0\Theta}.           &\nonumber
\end{eqnarray}
Here $\langle\dots\rangle$ means the trace operation, the tildes mean
that $\widetilde L_{\mu}$ and $\widetilde R_{\mu}$ are determined
without the singlet fields $L^0_{\mu}$ and $R^0_{\mu}$ (i.e.,
$\widetilde L_{\mu}^0 = \widetilde R_{\mu}^0 = 0$). As a result,
$\nabla_{\mu} U$ transforms like $U$, which is invariant under
U$(1)_L \times $U$(1)_R$. The same property is also relevant for
$M_{\Theta}$. The singlet external fields $L^0_{\mu}$ and $R^0_{\mu}$
are both collected in the last term in (9), $\nabla_{\!\mu}\Theta$ is
the chiral invariant derivative of $\Theta$. The quantities $F,B,H$
in (9) are the low-energy constants. Their physical significance may
be established basing on the property that at the leading order $p^2$
the quantum loops do not contribute in the effective theory
\cite{G-L-2}. Therefore the generating functional is equal to the
classical action,
\begin{equation}
L_{eff}(U;V,A,S,P,\Theta) = 
\int d^4 x {\cal L}_{eff}(U;V,A,S,P,\Theta),
\end{equation}
evaluated at the solution to the equations of motion for $U$. Owing
to this property one can show that $F$ is the decay constant of the
axial-vector octet quark current, $BF^2$ is the quark condensat (with
the opposite sign) in the chiral limit, $H$ describes the low-energy
asymptotic of the two-point Green function of the axial-vector
singlet quark current.

It is easy to see that due to (3) and (6) the lagrangian (9) and 
generating functional (10) are both chiral invariant. Moreover, they
are RG invariant as well. Really, owing to \cite{Shore} and the above
QCD description of the low-energy constants one can find that $F$ is
RG invariant, whereas constants $B$ and $H$ are renormalized as
\begin{equation}
B \, = \, Z^{-1}_{m} B_R,   \qquad  H \, =  \, Z^{-2}H_R.
\end{equation}
Since the external fields are renormalized by (7), the RG invariance
of lagrangian (9) is observed if matrix $U$ is RG invariant. However,
the latter property takes place due to the equations of motion.  So,
the RG invariance is really observed. Notice, since $U$ and $F$ are
both RG invariant, the interpolating fields $\eta^{\alpha}$ ($\alpha
= 1,\dots 8$) must be RG invariant, too, as it should be in a
consistent effective theory.

Now let allow for the presence of singlet states in the effective
theory. A way to correspondingly generalize the chiral effective
lagrangian is through the shift of the low-energy constants to
invariant functions \cite{G-L-2} which would describe the dependence
on the singlet-state interpolating fields. Besides, one must add the
necessary kinetic terms in order to describe the spectrum of the
singlet states. Of course, in this way one cannot distinguish between
different singlet states, and one must add some extra considerations
to do that. This problem will be intensively discussed below. For the
present, we only notice one special case of singlet state. It is the
case of the lowest singlet quarkic state $\eta^0$ which interpolating
field must transform by a shift under U(1)$_{A}$:
\begin{equation}
\eta^0 \, \to \, \eta^0 + F_0\omega^0_5.
\end{equation}
Condition (12) was first imposed in \cite{W1} in order to resolve in
terms of the effective theory the paradox between the
$\theta$-dependence of the vacuum and the large-$N_c$ behavior of QCD
with massless quarks. The quantity $F_0$ is a dimensional parameter.
In the limit of large $N_c$ it must coincide with $F$ \cite{W1}, but
its value remains unknown with $N_c$ finite. Due to (12) and (6), the
combination $\eta^0 + F_0\Theta$ is completely chiral invariant. So,
this very combination should be placed into the lagrangian, but not
$\eta^0$ itself.

This important observation was made, and the corresponding
generalization of the chiral effective lagrangian that involves
$\eta^0$ was proposed \cite{G-L-2}. However, the physical
significance of $F_0$, especially that in terms of QCD parameters, is
not clarified yet. (Usually $F_0$ is equated to $F$ without
discussions.) This gap, of course, should necessarily be filled.
Besides, there is another problem while involving $\eta^0$.  It is
the necessity to observe the RG symmetry inspired by QCD.
Previously, it was not taken into consideration in the singlet
channel. As a result, the RG symmetry was lost. Consequently, for
instance, the coupling of the singlet axial-vector current to $\eta'$
was described wrong. This gap should be filled, too. Our nearest task
is to solve these problems.

First, let us find the correct generalization of ${\cal L}_{eff}$
which would involve $\eta^0$ and satisfy both the chiral and RG
symmetry. Again, we will work up to order $p^2$ of the chiral
expansion. In general case ${\cal L}_{eff}$ may be represented in the
form
\begin{equation}
{\cal L}_{ eff} = 
{\cal L}^{(0)} + {\cal L}^{(kin)} + {\cal L}^{(mass)},
\end{equation}
where ${\cal L}^{(0)}$ involves $\eta^0$ only, without contributions
of the octet interpolating fields. ${\cal L}^{(kin)}$ and ${\cal
L}^{(mass)}$ in (13) involve both $\eta^0$ and the octet
interpolating fields. It is convenient to take ${\cal L}^{(0)}$ in
the form with the explicitly extracted quadratic terms:
\begin{eqnarray}
& \!\!\!\!\! {\cal L}^{(0)}  = \,{1 \over 2} 
                   \nabla_{\!\mu}\eta^0 \,\nabla^{\mu}\eta^0 - \,
{1 \over 2} M^2_0 (\eta^0 + F_0\Theta)^2  &             \\[1mm]
& \quad\qquad\; + \>\,H_0 \,\nabla_{\!\mu}\eta^0 \,\nabla^{\mu}\Theta
+ {1 \over 2} H \,\nabla_{\!\mu}\Theta \nabla^{\mu}\Theta 
                 + {\cal L}^{(0)}_{int}.  &             \nonumber
\end{eqnarray}
Here 
\begin{equation}
\nabla_{\!\mu}\eta^0 = \partial_{\mu}(\eta^0 + F_0\Theta)
\end{equation}
is the chiral-invariant derivative of $\eta^0$. Note, we have defined
it differently as compared with \cite{G-L-2} where
$\nabla_{\!\mu}\eta^0$ was defined using $A^0_{\mu}$ instead of
$\partial_{\mu}\Theta$. Although from the point of view of the chiral
symmetry both definitions are equivalent, our one (15) is preferable
from the point of view of RG symmetry (see below). $H_0$ in (14) is a
new dimensional low-energy constant (its physical significance will
be discussed latter). $M_0$ is the mass of $\eta^0$ in the chiral
limit (it may be related to the topological susceptibility of gluons
in QCD without quarks \cite{W1}). The term ${\cal L}^{(0)}_{int}$ in
(14) describes the $\eta^0$ self-interaction and its accompanying
interaction with external fields. This term is irrelevant, however,
when only the mixing and decays of $\eta^0$ are subjects of
consideration. The `kinetic' and the `mass' terms in (13) are as
follows
\begin{eqnarray}
& \!\!\!\!\!\! \!\!\!\!\!\!\!\!\!\!\!\!\!\!\!\!
 {\cal L}^{(kin)} \>  = \, {1 \over 4} F^2
\upsilon_1\, \langle {\nabla}_{\!\mu}U
	{\nabla}^{\mu}U^{\dagger}\rangle , & \\[1mm]
& {\cal L}^{(mass)} \!  = \, {1 \over 2} B F^2 \>
             \langle  M_{\Theta} \upsilon_2^{\ast} U^{\dagger} + 
M_{\Theta}^{\dagger} \upsilon_2 U \, \rangle . &
\end{eqnarray}
Here constants $F$ and $B$ are the same as in (9), while $\upsilon_1$
and $\upsilon_2$ are invariant functions on $\eta^0 + F_0\Theta$.
Their normalization is chosen so, that their expansions in the powers
of fields start with 1. Due to the parity and charge-conjugation
invariance $\upsilon_1$ must be real and even, whereas $\upsilon_2$
may be complex and $\upsilon_2^{\ast}(\mbox{x}) = \upsilon_2
(-\mbox{x})$.\footnote{ Notice, sometimes $\upsilon_2$ is defined
with the extracted factor $\exp \,i\lambda^0 (\eta^0/F_0+\Theta)$.
Then, the nonet matrix $\Sigma = U \exp (i\lambda^0 \eta^0/F_0)$ may
be inserted into (17) instead of $U$, and the external-field
combination $M$ instead of $M_{\Theta}$ \cite{G-L-2}. However,
actually, there are no physical reasons to do this modification since
only a few degrees of $\eta^0$ are really significant in (17).}

Now let us discuss RG properties of lagrangian (13). The crucial
question is RG property of the constant $F_0$. Strictly speaking, one
cannot solve this question until the representation of $F_0$ is found
in terms of QCD parameters. Nevertheless, basing on the common sense,
one may assume that $F_0$ is RG invariant, since its value, according
to (12), is related to the normalization of the interpolating field
$\eta^0$. Assuming this property (for the strict proof see the next
section) and taking into account (7), one gets that $\eta^0 +
F_0\Theta$ is RG invariant. The next important observation is that
the dependance on $A^0_{\mu}$ in lagrangian (13) is only realized
through $\nabla_{\!\mu}\Theta$. As a result and, again, due to (7),
RG invariance of ${\cal L}_{ eff}$ is observed if there is the
following renormalization rule, in addition to (11),
\begin{equation}
H_0 = Z^{-1}H_{0\,R} .
\end{equation}
Actually, (18) follows from (8) and the fact that parameter $H_0$ is
the decay constant of the axial-vector singlet quark current. The
simplest way to verify that is to examine this current in the
effective theory. In accordance with its natural definition as the
variational derivative of the action we have 
\begin{equation}
{\cal J}^0_{\mu 5} \> \equiv \>  
\frac{\delta L_{eff}}{\delta A^{0\mu}}  \> = \> - \>  
\frac{\partial{\cal L}_{eff}}{\partial (\nabla^{\mu}\Theta)} \> = \>
- \> H_0\,\nabla_{\!\mu} \eta^0 + H\,\nabla_{\!\mu} \Theta + \dots
\end{equation}
Here dots mean irrelevant higher-order terms of the expansion in the
powers of fields. The straightforward consequence of (19) is
$<0|{\cal J}^0_{\mu 5}|\eta^0\!> = -i\, H_0 p_{\mu}$, which was to be
proved. A more correct proof is based on the analysis of two-point
Green function $\delta^2 W/\delta A^0_{\mu}\delta A^0_{\nu}$. When it
is considered in QCD, its residue over the pole of $\eta^0$ is equal
to the matrix element $<0|\mbox{\sf J}^0_{\mu 5}|\eta^0\!>$ in
square. In the effective theory the direct calculation leads to the
squared $H_0 p_{\mu}$ for this quantity. So, $H_0$ is really the
decay constant of the current $\mbox{\sf J}^0_{\mu 5}$.

It is interesting to compare (19) with a similar expression for the
Noether U$(1)_A$ current. In accordance with (3) and (12) it has the
form
\begin{equation}
\Im^0_{\mu 5} \> \equiv \> 
F_0 \frac{\partial{\cal L}_{eff}}{\partial(\partial^{\mu}\eta^0)} 
\> = \>
F_0 \frac{\partial{\cal L}_{eff}}{\partial(\nabla^{\mu}\eta^0)} 
\> = \>
F_0\nabla_{\!\mu} \eta^0 + \dots
\end{equation}
Here $F_0$ plays a similar role as $H_0$ in (19) but just in the
Noether current. So, $F_0$ has the meaning of the `decay' constant of
the Noether current in the effective theory. However, this meaning is
valid no longer in QCD where the Noether U$(1)_A$ current is
$\mbox{\sf J}^0_{\mu 5} - \mbox{\sf K}_{\mu}$ with $\mbox{\sf
K}_{\mu}$ is the pure gauge-field current, $\partial_{\mu}\mbox{\sf
K}^{\mu} = \mbox{\sf Q}$. Really, in QCD the Noether U$(1)_A$ current
is conserved. Therefore, in the chiral limit it cannot generate any
massive PS state, including $\eta^0$ one, since the divergence of the
current is zero. On the contrary, in the effective theory the Noether
U$(1)_A$ current is not conserved.  Therefore, it is able to generate
$\eta^0$, which is explicitly exhibited in (20). Let us note, that in
fact the U$(1)_A$ symmetry is broken not only in the effective theory
but in QCD, too. However, the nature of the breaking is different in
both theories. Indeed, in the effective theory the symmetry is
explicitly broken due to non-invariance of the lagrangian (with the
external fields fixed or switched-off). In QCD the lagrangian is
quasi-invariant under global U$(1)_A$, i.e. it transforms on a total
divergence (therefore there is the conserved Noether current), and
the symmetry is broken due to nonperturbative effects in
$\theta$-vacuum \cite{W1,t'H}.

The difference between Noether U$(1)_A$ currents in QCD and in the
effective theory may be related to the property that the
effective-theory analog of the gluon anomaly is not a divergence.
Indeed, in the effective theory the `gluon anomaly' operator is
\begin{equation}
{\cal Q} \> \equiv \> 
\frac{\delta L_{eff}}{\delta\Theta} \> = \>
\left(F_0 \frac{\delta{\cal L}_{eff}}{\delta\eta^0}\right)
 + \lambda^0 \left( S^a \frac{\partial{\cal L}_{eff}}{\partial P^a} 
 - P^a \frac{\partial{\cal L}_{eff}}{\partial S^a} \right)
- \partial_{\mu} 
\frac{\partial{\cal L}_{eff}}{\partial\nabla_{\mu}\Theta}.
\end{equation}
Here the first term in the r.h.s. is neither a divergence and nor a
zero when the equation of motion for $\eta^0$ is not imposed. So, the
same property must be peculiar to ${\cal Q}$, whereas in QCD
$\mbox{\sf Q} \equiv \partial_{\mu}\mbox{\sf K}^{\mu}$. Actually,
this difference is natural, because ${\cal Q}$ is only able to
describe the observable degrees of freedom, and therefore it cannot
`feel' the presence of the gauge-variant current $\mbox{\sf K}^{\mu}$
in QCD. Nevertheless, on the equations of motion both ${\cal Q}$ and
{\sf Q} become equivalent. Indeed, from (21) and (19) there follows
the anomalous Ward identity for ${\cal Q}$ which is the same as that
for {\sf Q}. Owing to (7) the RG property (8) is satisfied for ${\cal
Q}$ as well. Notice, with the presence of quarks {\sf Q} is
completely gauge-invariant operator on the equations of motion (e.g.,
due to the anomalous Ward identity).

Remember, the operator {\sf Q} is able to generate PS glueballs.
Besides, it is able to generate the quarkic singlet PS state
$\eta^0$, but only in the next-to-leading order in the large-$N_c$
\cite{W1,W2}. In the effective theory this property is reproducible,
too. Really, with the external fields switched off and the equations
of motion taken into account, (21) reads
\begin{equation}
{\cal Q} \> = \> H_0 M_0^2 \eta^0 \> + \> 
\mu_0 \,C_0 \eta^0 \> + \> \mu_8 \,C_8 \eta^8 \> + \> \dots
\end{equation}
Here dots stand for multiparticle contributions, the quantities
$\mu_0$ and $\mu_8$ are the RG invariant combinations
$B(m_u+m_d+m_s)$ and $B (m_u+m_d-2m_s)$. Parameter $C_{0}$ in (22)
originates from the power-field expansion of $\upsilon_2$,
$C_8=\frac{\sqrt{2}}{3}F$. Since $H_0 \sim N_c^{1/2}$ and $M_0^2 \sim
N_c^{-1}$ at large $N_c$, then it follows from (22) that in the limit
of the massless quarks ${\cal Q} \sim N_c^{-1/2}$, whereas the
correct behavior is ${\cal Q} \sim N_c^0$. To reproduce this correct
behavior one has to take into consideration another singlet state,
$\eta^G$, which is gluonic by its origin. The corresponding
generalization of the lagrangian is obvious: one should allow for the
$\eta^G$-dependence in $\upsilon_{1,2}$ and the necessary terms in
${\cal L}^{(0)}$. The latter one now reads
\begin{eqnarray}
& \!\!\!\!\!\!\!\!\!\!\!\!\!\!\!\!\!\!\!\!\!\!\!\!\!\!\!
  \!\!\!\!\!\!\!\!\!\!\!\!\!\!\!\!\!\!\!\!\!\!\!\!\!\!
  {\cal L}^{(0)}  = 
  {1 \over 2}\nabla_{\!\mu}\eta^0 \,\nabla^{\mu}\eta^0 -
  {1 \over 2} M^2_0 (\eta^0 + F_0\Theta)^2         &        \\[1.5mm]
& \!\!\!\!\!\!\!\!\!\!\!\!
 + \> {1 \over 2}\partial_{\mu}\eta^G \,\partial^{\mu}\eta^G - 
      {1 \over 2} M^2_{G}(\eta^G)^2 - q\,(\eta^0 + F_0\Theta)\eta^G
&                                                  \nonumber\\[0.5mm]
& \qquad
 + \> H_0\,\nabla_{\!\mu}\eta^0 \, \nabla^{\mu}\Theta
 + H_{G} \partial_{\mu}\eta^G \nabla^{\mu}\Theta 
 + {1 \over 2} H \,\nabla_{\!\mu}\Theta \nabla^{\mu}\Theta 
 + {\cal L}^{(0)}_{int} &.                         \nonumber
\end{eqnarray}
Here $q$ is a new parameter that describes the mixing between
$\eta^0$ and $\eta^G$, $M_G$ is a mass parameter for $\eta^G$, $H_G$
is another decay constant of the current $\mbox{\sf J}^0_{\mu 5}$.
The gluonic nature of $\eta^G$ is developed in the specific
large-$N_c$ behavior of its parameters. Namely, since $\eta^G$ is a
gluonic state, then $H_{G} \sim N_c^0$ and $M_{G}^2 \sim N_c^0$.
Besides, since $\eta^0$ is a quarkic state, then $q \sim N_c^{-1/2}$.
(See \cite{G-L-2} and \cite{W2} for the way to show that.) So, now
\begin{eqnarray}
{\cal Q} & = & (H_{G} M_G^2 + H_0 q) \,\eta^G \> + \> 
(H_0 M_0^2 + H_{G} q) \,\eta^0                              \\[0.5mm]
         & + & \mu_0 \,C_0 \eta^0 \> + \> 
               \mu_8 \,C_8 \eta^8 \> + \> \dots             \nonumber
\end{eqnarray}
has the correct behavior at large $N_c$, which is caused by the
presence of $\eta^G$.

It is important to note that in (23) we did not introduce the
kinetic-mixing term $\nabla_{\!\mu}\eta^0\partial^{\mu}\eta^G$
because we considered $\eta^0$ and $\eta^G$ to be independent
canonical variables. The latter property follows from the condition
that $\eta^0$ and $\eta^G$ must describe quite different degrees of
freedom. (Namely, the quarkic and gluonic ones, which we assume to
exist on the projection to interpolating fields in QCD. This is quite
a general assumption and we do not consider it as a model-dependent
one. Notice, an equivalent assumption reads that there exist
glueballs and meson quarkic states in QCD.)

The RG properties of the parameters of $\eta^G$ in lagrangian (23)
may be established analogously to those of $\eta^0$. In this way, $q$
and $M_G$ must be RG invariant, like $F_0$ does, in order to provide
the lagrangian with RG invariance. Since $H_G$ is the decay constant
of the axial-vector singlet quark current, its renormalization rule
must coincide with that of $H_0$. An important consequence of these
RG properties is the scale-dependent mixing of the quarkic and
gluonic contributions to the gluon anomaly operator.  Looking at (24)
we obtain this property due to RG non-invariance of the first two
terms and invariance of the last two terms. In QCD a similar RG
behavior of the gluon anomaly was established in \cite{Shore} and
discussed in detail in \cite{N1}. This behavior means that in QCD
with quarks the gluon anomaly operator has no pure gluonic nature,
but it rather has a mixed nature. Actually, the mixed nature is
peculiar to other QCD composite operators which mix under RG.
Especially it becomes clear when the operators are considered on the
equations of motion, since even the fundamental fields of quarks and
gluons carry the mixed degrees of freedom on the equations of motion.

So, the composite operators appear to be not quite suitable variables
for description of singlet states. The preferable variables appear to
be the quarkic and gluonic interpolating fields because they are RG
invariant and available for a direct description of the quarkic and
gluonic degrees of freedom of the observable states. In the framework
of QCD these interpolating fields may be introduced in rather
indirect manner, basing on the composite operators as the initial
objects \cite{N1}. In the framework of the effective theory these
interpolating fields are introduced as the fundamental objects. This
peculiarity shows a certain advantage in describing singlet states in
the framework of the effective theory.

Now, so long as the lowest quarkic and gluonic states have been
introduced, one can make the next step and introduce other singlet
states. Assuming that each new singlet state presents its own unique
degree of freedom, one has to assume that its interpolating field
must be independent canonical variable. In what follows we will
reserve the symbol $\eta^{\kappa}$ for any extra singlet PS state if
it does not coincide with $\eta^0$ and we are not interested in its
nature. Special attention will be payed to the gluonic ground-state
$\eta^G$ and an excitation state over $\eta^0$, which will be
designated by the symbol $\widetilde\eta^{0}$. The difference between
$\eta^G$ and $\widetilde\eta^{0}$ may be detected in their
large-$N_c$ behavior.  For instance, parameter $\widetilde q$, which
describes the mixing between $\eta^0$ and $\widetilde\eta^{0}$, must
behave as $\widetilde q \sim N_c^{-1}$ whereas $q \sim N_c^{-1/2}$.
In principle, in this way one may unambiguously distinguish between
the states $\eta^G$ and $\widetilde\eta^{0}$. However, the
large-$N_c$ approach alone will hardly be useful to obtain any
significant phenomenological result.  A more promising way seems to
be in searching for more strong constraints for singlet-state
contributions to the lagrangian.

\section{ Singlet-state constraints }

Let us return to the problem of the physical significance of $F_0$.
We have seen that $F_0$ does not contribute to the axial-vector
singlet quark current and to the gluon anomaly operator. Therefore
$F_0$ cannot be expressed in terms of QCD Green functions with
corresponding legs. Let us examine now the pseudoscalar quark
current. First, one must define the power-field expansion of the
invariant function $\upsilon_2$,
\begin{equation}
\upsilon_2 \> = \> 
1\,+ \,i\lambda^0 \left\{ b_0 (\eta^0 /F_0 + \Theta) + 
                         \sum_{\kappa} b_{\kappa}\eta^{\kappa}/F
                         \right\} + \dots
\end{equation}
Here $\lambda^0 = \sqrt{2/3}$ is the numerical factor, $b_0$ and
$b_{\kappa}$ are the parameters of the linear term of the expansion,
dots stand for the higher-order terms of the expansion. In view of
(25) and (17) the PS quark current in the effective theory is
\begin{equation}
{\cal J}^0_5  \equiv  
-\> \delta L_{eff}/\delta P^0 = \,
-\,BF^2 \left( b_0  \eta^0 / F_0 + 
              \sum_{\kappa} b_{\kappa}\eta^{\kappa}/F
              \right) + \dots
\end{equation} 
Here in the r.h.s the sources are switched off. Now let us take into
account the equality $<0|{\cal J}^0_5|\eta^0\!> \, = \, <0|\mbox{\sf
J}^0_5|\eta^0\!>$ which follows from the equality of the residues
over the $\eta^0$-pole in Green function $\delta^2 W/\delta P^0\delta
P^0$ in the effective theory and in QCD. Then, owing to $BF^2 =
-<\!\bar u u\!>_{_{\!0}}$, where $<\!\bar u u\!>_{_{\!0}}$ is the
chiral quark condensat ($<\!\bar u u\!>_{_{\!0}}\> = <\!\bar d
d\!>_{_{\!0}} \> = \> <\!\bar s s \!>_{_{\!0}}$), one may obtain from
(26) the relation
\begin{equation}
\frac{F_0}{b_0} \> = \>
\frac{<\!\bar u u\!>_{_{\!0}}}{<0|\mbox{\sf J}^0_5|\eta^0>}.
\end{equation} 
Owing to (27) the physical significance of the ratio $F_0/b_0$ is
clear: it presents the coefficient which should be extracted with the
inverse quark condensat from the QCD composite operator $\mbox{\sf
J}^0_5$ in order to obtain the canonically normalized interpolating
$\eta^0$ field on the mass shell of $\eta^0$. Note, a similar
expression for this coefficient was obtained in \cite{S-V} where also
the RG invariance property of this coefficient was discussed.

However, the physical significance of the parameter $F_0$ proper
still remains unclear. In this connection, let us also examine the
U(1)$_A$ transformation property of the current $\mbox{\sf J}^0_5 =
i\bar q \gamma_5 \lambda^0/2 \,q$. To this end let perform rotation
$q \to \exp (-i\gamma_5 \omega^0_5 \lambda^0 /2)q$, $\bar q \to \bar
q \exp (-i\gamma_5 \omega^0_5 \lambda^0 /2)$. Then obtain with
infinitesimal $\omega^0_5$
\begin{eqnarray}
& \mbox{\sf J}^0_5 \> \rightarrow \> \mbox{\sf J}^0_5 + 
  \mbox{\sf J}^0 \lambda^0 \omega^0_5.
\end{eqnarray}
Here $\mbox{\sf J}^0 = \bar q \lambda^0/2q$. Now let us take into
consideration the expansions in the powers of interpolating fields of
the currents $\mbox{\sf J}^0_5$ and $\mbox{\sf J}^0$:
\begin{eqnarray}
\mbox{\sf J}^0_5 & = &
      <\!0 | \mbox{\sf J}^0_5 | \eta^0\!> \,\eta^0 \> + \> 
       \sum_{\kappa } <\!0 | \mbox{\sf J}^0_5 | \eta^{\kappa }\!> 
       \,\eta^{\kappa } \, +  \dots                                \\
\mbox{\sf J}^0   & = & 
      (\lambda^0)^{-1} <\!\bar u u\!>_{_{\!0}} \, +  \dots 
\end{eqnarray}
Here in (29) $\eta^0$ and $\eta^{\kappa }$ are canonically normalized
interpolating fields for $|\eta^0\!>$ and $|\eta^{\kappa }\!>$, dots
stand for multiparticle contributions. In (30) the first term in the
r.h.s. is the v.e.v. of the current $\mbox{\sf J}^0$, dots stand for
the irrelevant scalar-particle contributions and multiparticle
contributions. Substitute (29) and (30) into (28).  Then, assuming
the chiral-invariance of $\eta^{\kappa}$, we obtain that $\eta^0$
should transform on a shift:
\begin{equation}
\eta^0 \> \rightarrow \> \eta^0 \, + \,
\frac{<\!\bar u u\!>_{_{\!0}}}{<0|\mbox{\sf J}^0_5|\eta^0>}
\,\omega^0_5.
\end{equation}
Note, this $\eta^0$ has been defined directly in QCD. So, if one
identifies this $\eta^0$ with that in the effective theory, then one
confirms condition (12). Moreover, its origin becomes clear: it is
the consequence just of the nonzero v.e.v. of the current $\mbox{\sf
J}^0$ which is the chiral partner of $\mbox{\sf J}^0_5$.  Comparing
(12) with (31), one can also deduce the QCD representation of $F_0$.
It turns out to be exactly the r.h.s. of (27). So, since the
normalization of $\eta^0$ is the same everywhere (the canonical one),
there is consistency condition on the parameter $b_0$,
\begin{equation}
b_0 \> = \> 1.
\end{equation}

In view of (32) and (27) the physical significance of $F_0$ is found
to be the same which was before for the ratio $F_0/b_0$.
Simultaneously, both in QCD and in the effective theory parameter
$F_0$ governs the U(1)$_A$ transformation property of the
interpolating $\eta^0$ field.

Actually, consistency condition (32) is the universal one since it is
valid not matter what number of extra singlet states has been
involved/integrated out. So, condition (32) is able to constrain the
contributions of other singlet states to the chiral effective
lagrangian. Indeed, if any $\eta^{\kappa }$ contributes linearly to
expansion (25), i.e.  if $b_{\kappa } \not= 0$, then it should be
$q_{\kappa } = 0$, that is $\eta^{\kappa }$ does not mix with
$\eta^0$ in ${\cal L}^{(0)}$.  (Otherwise, after $\eta^{\kappa }$ is
integrated out there will appear an extra dependence on $\eta^0$ in
(25) which is caused by the former mixing, and this extra dependence
will break condition (32).) If, on the contrary, $b_{\kappa } = 0$,
then it well may be $q_{\kappa } \not= 0$. Moreover, in accordance
with the Weinberg `theorem' \cite{Weinberg} if it is allowed
$q_{\kappa } \not= 0$, then it should be $q_{\kappa } \not= 0$.

So, we have obtained a strict result that $\eta^{\kappa }$ cannot
contribute simultaneously to the linear term of the expansion of
$\upsilon_2$ and to the mixing $\eta^0\eta^{\kappa}$-term in
lagrangian ${\cal L}^{(0)}$. Therefore, there are two quite different
ways to involve an extra singlet state to the effective theory. This
fact inspires an idea that each way corresponds to some specific kind
of the singlet state. In the reality that indeed takes place. To show
this let us consider the limit of the massless quarks, when the octet
states become the Goldstone bosons but $\eta^0$ and $\eta^{\kappa}$
remain massive states. Then the condition $q_{\kappa }=0$ means that
$\eta^{\kappa }$ cannot be converted into the ground-state $\eta^0$
without the emission of some number of Goldstone bosons. However,
this behavior is peculiar exactly to excited states, since when there
is no mass (energy) gap massless strong-interacting particles {\it
should necessarily be emitted} in course of any transformation of an
excited state. (One may consider this property as an independent
definition of excited states in the framework of the effective
theory.) So, the condition $q_{\kappa }=0$ may be considered as the
necessary condition for excited states. Another case, when
$\eta^{\kappa }$ does not contribute to the linear term in
$\upsilon_2$ ($b_{\kappa }=0$), is quite natural for glueballs.
Indeed, since gluons do not distinguish the quark flavors, a pure
glueball {\it cannot contribute directly} (through the vertex) to any
process which breaks the flavor symmetry.  Therefore, the glueball
interpolating field cannot appear in lagrangian ${\cal L}^{(mass)}$
which explicitly breaks the flavor symmetry.\footnote{ It is
interesting to note, that, consequently, glueballs cannot contribute
to scalar and pseudoscalar quark currents ${\cal J}^0$ and ${\cal
J}^0_5$ ($\mbox{\sf J}^0$ and $\mbox{\sf J}^0_5$). Besides, the pure
PS glueball $\eta^G$ does not contribute to ${\cal Q}$ ({\sf Q})
through quark-mass-dependent terms (see Eq.  (24); cf. \cite{N1}).}
The latter property should be regarded as the necessary condition for
glueballs.

The results of the discussion, which concern the states $\eta^G$ and
$\widetilde \eta^0$, are summarized in the Table,

\medskip

\begin{displaymath}
\begin{array}{l|c|c}  \hline
& & \\[-2.5mm]
\qquad {\rm The \ kind \ of \ the \ extra }
&\quad {\rm contributes \ into} \quad
& \quad {\rm mixes \ with \ } \eta^0 \quad \\ 
\qquad {\rm singlet \ PS \ state \ \ } 
(\eta^{\kappa }) \qquad
&  \upsilon_2 \,{\rm\ in \ }\,{\cal L}^{(mass)} 
& {\rm in \ } \, {\cal L}^{(0)}            \\[-2.5mm]
& & \\ \hline
& & \\[-2.5mm]
\qquad \mbox{ground-state glueball \ } (\eta^G) \quad
& \,\mbox{no } \; (b_G = 0) & \mbox{ yes } (q \not= 0)        \\
\qquad {\rm excited \ state \ \ } (\widetilde\eta^0)
& \mbox{yes } (\widetilde b_0 \not= 0) & \mbox{ no }\, 
                                     (\widetilde q = 0)  \\[-2.5mm]
& & \\  \hline
\end{array}
\end{displaymath}
\\
\noindent
In addition to the Table, it should be noted that by the above
reasons the mixing between $\eta^G$ and $\widetilde\eta^0$ is
suppressed in ${\cal L}^{(0)}$. Besides, any excitation of the PS
glueball ($\widetilde \eta^G$) does not contribute both to the linear
term of $\upsilon_2$ ($\widetilde b_G = 0$) and to the mixing
$\widetilde \eta^G \eta^0$-term in ${\cal L}^{(0)}$ ($\widetilde q_G
= 0$). A analogous analysis may be extended to any other singlet
state.

\section{ Mixing }

The results of the previous section are most important for
investigation of the spectrum of singlet states. Let us discuss
briefly this question making the accent on the mixing phenomenon. The
simplest case is when only the ground-states $\eta^0$ and $\eta^G$
are involved as singlet states. Then, neglecting the isotopic
symmetry breaking ($m_u = m_d \not= m_s$), one has three iso-singlet
mixing states: $\eta^8$, $\eta^0$, and $\eta^G$. In virtue of the
above Table and (13), (17), (23), the squared mass matrix in the
basis of these states is
\begin{equation}
{\cal M}^2 = \left( \begin{array}{lll}
d_8            & r  a                 &  0            \\
               & M_0^2 + \beta_0 d_0  &  q            \\
\mbox{ symm.}  &                      &  M_G^2        \\
\end{array} \right) .
\end{equation}
Here
\begin{eqnarray}
& d_8 = \, {1 \over 3} (4 M_K^2 - M_{\pi}^2), \quad
  d_0 = \, {1 \over 3} (2 M_K^2\!+\!M_{\pi}^2), \quad
  a   = \, {2\sqrt{2} \over 3} (M_{\pi}^2 - M_K^2) , & \nonumber
\end{eqnarray}
$M_{\pi}$ and $M_K$ are pion and kaon masses, $r = F/F_0$, and
$\beta_0$ is the parameter of the quadratic $(\eta^0)^2$-term in the
expansion of $\upsilon_2$. Matrix ${\cal M}^2$ may be diagonalized by
the orthogonal rotation matrix
\begin{equation}
{\cal O} = \left[{\cal O}\right]^j_n = \left( \begin{array}{lll}
\>\> c_2 c_3                &\quad s_2     &\>\>\> c_2 s_3         \\
\>\>s_1 s_3 - c_1 s_2 c_3 &\quad c_1 c_2 &\> -s_1 c_3 - c_1 s_2 s_3\\
-c_1 s_3 - s_1 s_2 c_3 &\quad s_1 c_2 &\>\>\>c_1 c_3 - s_1 s_2 s_3 \\
\end{array} \right) .
\end{equation}
Here $c_i = \cos\theta_i$, $s_i = \sin\theta_i$; $\theta_1 =
\theta_{0\mbox{-}G}$, $\theta_2 = \theta_{8\mbox{-}0}$, $\theta_3 =
\theta_{8\mbox{-}G}$. The row index $j$ and the column index $n$ run
the values $ j = 8,0,G$ and $n = \eta,\eta',\eta^{_{Gl}}$.

One can obtain the following relations between the angles $\theta_i$
and the parameters of the matrix ${\cal M}^2$,
\begin{eqnarray}
& \tan\theta_1 & = \>\> 
{\cal O}^G_{\eta'}/{\cal O}^0_{\eta'} \>= \>
-\,\frac{q}{M^2_{G} - M_0^2 - \beta_0 d_0},                      \\
& \tan\theta_2 & = \>\> 
c_1 \cdot {\cal O}^8_{\eta'}/{\cal O}^0_{\eta'}
\>= \>\frac{a c_1}{ M_0^2 + \beta_0 d_0 -  d_8},                 \\
& \tan\theta_3 & = \,-\,\frac{s_1 + c_1 s_2 t_3}{c_2} \cdot 
{\cal O}^8_{\eta^{_{Gl}}}/{\cal O}^0_{\eta^{_{Gl}}}
\>= \, - \,\frac{a s_1}{ M^2_{G}}.
\end{eqnarray}
Here in each formula the second equality displays the leading term
of the combined chiral and large-$N_c$ expansion. Remember,
$M_{\pi,K}^2 = {\mbox O}(p^2,1)$, $M_0^2 = {\mbox O}(1,N_c^{-1})$,
$M_G^2 = {\mbox O}(1,1)$, $q = {\mbox O}(1,N_c^{-1/2})$.
Consequently, $M_{\eta}^2 = {\mbox O}(p^2,1)$, $M_{\eta'}^2 ={\mbox
O}(p^2,N_c^{-1})$, $M_{\eta^{_{Gl}}}^2 = {\mbox O}(1,1)$.

In what follows, we will use a common parameter $\varepsilon$ of the
combined expansion. We choose it to be of the order ${\mbox O}(p)$ in
the sense of the chiral expansion and of the order of some negative
power of $N_c$ in the sense of the large-$N_c$ expansion,
\begin{equation}
\varepsilon \> = \> {\mbox O}(p) \> = \> 
{\mbox O}(N_c^{-\alpha}).
\end{equation}
Actually, one should consider $\alpha > 1/2$, since otherwise
$\eta^0$ could not be a heavy state as compared with pions and kaons.
The real value of $\alpha$ may be estimated like as follows. Let, in
accordance with (38), $M_{\eta}^2 \sim \varepsilon^2$, $M_{\eta'}^2
\sim \varepsilon^{1/\alpha}$, $M_{\eta^{_{Gl}}}^2 \sim 1$. Then one
can obtain $\alpha \simeq \frac{1}{2} \ln\left(M_{\eta^{_{Gl}}}^2 /
M_{\eta}^2 \right) / \ln\left(M_{\eta^{_{Gl}}}^2 / M_{\eta'}^2
\right)$. As a consequence, $\alpha =\,$0.9--1.1 in wide mass region
$M_{\eta^{_{Gl}}} = \,$(1.5--1.9)GeV. So, one may put approximately
$\alpha \simeq 1$. With this value of $\alpha$ from (35)--(37) we
have the hierarchy of the angles:
\begin{equation}
\theta_1 = \theta_{0\mbox{-}G} \sim \varepsilon^{1/2}, \quad\; 
\theta_2 = \theta_{8\mbox{-}0} \sim \varepsilon,       \quad\;
\theta_3 = \theta_{8\mbox{-}G} \sim \varepsilon^{5/2} .
\end{equation}
Note, the mixing $\eta^8 - \eta^G$ is the smallest one because it
arises only due to intermediate mixing with $\eta^0$.

Now let us consider the most interesting case when $\eta^G$ and
$\widetilde\eta^0$ are both involved together with $\eta^0$. The
squared mass matrix in the basis $\eta^8 - \eta^0 - \eta^G -
\widetilde\eta^0$ has the form\footnote{ Strictly speaking, involving
$\widetilde\eta^0$, one must also take into account
$\widetilde\eta^8$, where $\widetilde\eta^8$ is the eighth member of
the nonet of excited states. (A general way to include the octet
heavy states is discussed in \cite{E-G-P-DR}.) Then,
$\widetilde\eta^0$ and $\widetilde\eta^8$ must mix in lagrangian
${\cal L}^{(mass)}$ to produce two final states. However, one of
these final states will decouple practically from the further mixing
with $\eta^8$, $\eta^0$, $\eta^G$, since $m_{u,d}\ll m_s$. As a
result, only the other state is really significant. Since its
contribution to the mixing is quite similar to that of
$\widetilde\eta^0$, considered without $\widetilde\eta^8$, we will
neglect for simplicity the effect of the presence of
$\widetilde\eta^8$.}
\begin{equation}
\mbox{\bf M}^2 = \left( \begin{array}{ccccc}
  &             &  & \vdots  & \widetilde b_0 a        \\[-1mm]
  & {\cal M}^2  &  & \vdots  & \widetilde \beta_0 d_0  \\[-1.5mm]
  &             &  & \vdots  &  0                      \\[-3mm]
\ldots & \ldots & \ldots & \ldots & \ldots             \\[-2mm]
\widetilde b_0 a   & \widetilde \beta_0 d_0 & 0 & \vdots  & 
M_{\widetilde\eta^0}^2 \\
\end{array} \right) .
\end{equation}
Here ${\cal M}^2$ is given by (33). The zeros in (40) reflect the
absence of the direct mixing between $\eta^G$ and $\widetilde\eta^0$.
Parameter $\widetilde \beta_0$ describes the mixing $\eta^0 -
\widetilde\eta^0$ caused by lagrangian ${\cal L}^{(mass)}$ (the
$\eta^0\widetilde\eta^0$-term in the expansion of $\upsilon_2$).

The diagonalization of the matrix $\mbox{\bf M}^2$ may be performed
in two steps. First, one may diagonalize the block ${\cal M}^2$:
\begin{eqnarray}
\mbox{\bf M}^2 \rightarrow 
\mbox{\bf M}^2_{\cal O} & = & 
\left( \begin{array}{ccc}
 {\cal O} & \vdots &  0      \\[-2.5mm]
      \ldots       & \ldots & \ldots \\[-2mm]
       0           & \vdots &  1      \\
\end{array} \right)^{-1}
\left( \begin{array}{ccc}
 {\cal M}^{2} & \vdots & \delta{\cal M}^2  \\[-2.5mm]
       \ldots       & \ldots & \ldots                    \\[-2mm]
 (\delta{\cal M}^2)^{T}& \vdots & M_{\widetilde\eta^0}^2  \\
\end{array} \right) \,
\left( \begin{array}{ccc}
	  {\cal O}& \vdots &  0      \\[-2.5mm]
       \ldots       & \ldots & \ldots \\[-2mm]
	  0           & \vdots &  1      \\
\end{array} \right)
\nonumber \\
 & = &
\left( \begin{array}{ccc}
{\cal M}^{2}_{{\cal O}}&\vdots &\delta{\cal M}^2_{{\cal O}}\\[-2.5mm]
       \ldots       & \ldots & \ldots                    \\[-2mm]
 (\delta{\cal M}^2_{{\cal O}})^{T}& \vdots & M_{\widetilde\eta^0}^2\\
\end{array} \right) .
\end{eqnarray}
Here ${\cal M}^{2}_{{\cal O}} = {\rm diag}(M_1^2,M_2^2,M_3^2)$, and
$\delta{\cal M}^2$ is the upright block in (40), 
\begin{equation}
\delta{\cal M}^2_{{\cal O}} \> \equiv \>
{\cal O}^{T} \delta{\cal M}^2 \> = \> 
\sqrt{4/3} \, M_K^2 \, {\cal Y}.
\end{equation}
In (42) ${\cal Y} = \mbox{column}(Y_1,Y_2,Y_3)$, $Y_n =
-\sqrt{2/3}\,{\cal O}^8_{n}\,\widetilde b_0 + \sqrt{1/3}\,{\cal
O}^0_{n}\,\widetilde \beta_0 $, and we neglected the $u$ and $d$
quark contributions as compared with the $s$-ones (approximation
$M^2_{\pi} \ll M^2_K$). Matrix $\mbox{\bf M}^2_{\cal O}$ may be
diagonalized by the next transformation
\begin{equation}
\mbox{\bf M}^2_{\cal O} \rightarrow
\mbox{\bf M}^2_{\Omega_{\sigma}} \> = \> 
\left( \begin{array}{ccc}
       1            &\!\vdots\!& -\, \sigma \\[-2.5mm]
   \ldots           & \ldots &    \ldots  \\[-2mm]
\sigma^{T} &\!\vdots\!&   1                 \\
\end{array} \right) 
\left( \begin{array}{ccc}
{\cal M}^{2}_{\cal O} &\!\vdots\!& \delta{\cal M}^2_{\cal O}\\[-2.5mm]
       \ldots       & \ldots & \ldots                    \\[-2mm]
 (\delta{\cal M}^2_{\cal O})^{T}&\!\vdots\!& M_{\widetilde\eta^0}^2 \\
\end{array} \right) 
\left( \begin{array}{ccc}
       1              &\!\vdots\!& \sigma \\[-2mm]
\ldots              & \ldots & \ldots  \\[-2mm]
-\,\sigma^{T} &\!\vdots\!&   1        \\
\end{array} \right) 
\quad \qquad 
\end{equation}
\begin{displaymath}
= \left( \begin{array}{ccc}
{\cal M}^{2}_{{\cal O}}\!-\!
\delta{\cal M}^2_{\cal O}\sigma^{T}\!\!-\!
	   \sigma (\delta{\cal M}^2_{\cal O})^{T}\!+\! 
	   \sigma M_{\widetilde\eta^0}^2\sigma^{T}
  & \! \vdots \! & 
{\cal M}^{2}_{{\cal O}}\sigma\!-\!\sigma M_{\widetilde\eta^0}^2\!+\!
		    \delta{\cal M}^2_{\cal O}\!-\!
     \sigma (\delta{\cal M}^2_{\cal O})^{T}\sigma \\[-2.5mm] 
\ldots       & \ldots & \ldots                    \\[-2mm]
\sigma^{T}{\cal M}^{2}_{{\cal O}}\!-\!
		     M_{\widetilde\eta^0}^2 \sigma^{T}\!\!+\!
		     (\delta{\cal M}^2_{\cal O})^{T}\!\!-\!
                     \sigma^{T} \delta{\cal M}^2_{\cal O}\sigma^{T}
& \! \vdots \! & 
M_{\widetilde\eta^0}^2\!+\!(\delta{\cal M}^2_{\cal O})^{T}\!\!+\!
\sigma^{T} \delta{\cal M}^2_{\cal O}\!+\! 
\sigma^{T} {\cal M}^2_{\cal O}\sigma 
\end{array} \right)
\end{displaymath}
Here the transformation matrix approaches the orthogonal one if
$\sigma = \mbox{O}(p^2)$ (we do not consider temporarily the
large-$N_c$ expansion). Assuming this property, we obtain that matrix
$\mbox{\bf M}^2_{\Omega_{\sigma}}$ is diagonal at order $p^2$ if
\begin{equation}
{\cal M}^{2}_{{\cal O}}\sigma - \sigma M_{\widetilde\eta^0}^2 +
		    \delta{\cal M}^2_{\cal O} \> = \> 0 .
\end{equation}
From (44) and (42) it follows that
\begin{equation}
\sigma_n \> = \> \sqrt{4/3} \, 
\frac{M_K^2}{M_{\widetilde\eta^0}^2-M_n^2}\,{\cal Y}_n.
\end{equation}
In view of $M_K^2 = \mbox{O}(p^2)$, $M_{\widetilde\eta^0}^2 =
\mbox{O}(1)$, one can deduce from (45) that $\sigma$ is really of
order $p^2$. So, up to and including order $p^2$ of the chiral
expansion the total diagonalizing matrix is
\begin{equation}
\Omega_{\sigma} \> = \> 
\left( \begin{array}{ccc}
     {\cal O }     & \vdots &  {\cal O}\sigma  \\[-2.5mm]
      \ldots       & \ldots & \ldots           \\[-2mm]
 -\sigma^{T}       & \vdots &   1              \\
\end{array} \right) ,
\end{equation}
where $\mbox{\bf M}^2_{\Omega_{\sigma}} =
\Omega_{\sigma}^{-1}\mbox{\bf M}^2 \Omega_{\sigma}$. In (46) each
$\sigma_n$ may be understood as the angle describing the mixing
between $\widetilde\eta^0$ and $\eta^8, \eta^0, \eta^G$. Owing to
(34), (39) and $\widetilde b_0, \widetilde \beta_0
\sim N_c^0$ we have
\begin{equation}
\theta_{8\mbox{-}\widetilde 0} \,\simeq\, \sigma_1 
                               \,\sim\, \varepsilon^2, \quad\;
\theta_{0\mbox{-}\widetilde 0} \,\simeq\, \sigma_2 
                               \,\sim\, \varepsilon^2, \quad\;
\theta_{ G\mbox{-}\widetilde 0} \,\simeq\, \sigma_3
                                \,\sim\, \varepsilon^{5/2}.
\end{equation}
So, $\theta_{0\mbox{-}G}$ and $\theta_{0\mbox{-}8}$ remain to be the
main mixing angles with the behavior of $\varepsilon^{1/2}$ and
$\varepsilon$, respectively. The next angles turn out to be
$\theta_{8\mbox{-}\widetilde 0}$ and $\theta_{0\mbox{-}\widetilde
0}$, both of order $\varepsilon^2$. The angles $\theta_{
G\mbox{-}\widetilde 0}$ and $\theta_{ 8\mbox{-}G}$ are the smallest
ones because they arise non-directly only, through the intermediate
mixing with $\eta^0$ (namely, through the mixings $\eta^0 -
\widetilde\eta^0$ in ${\cal L}^{(mass)}$ and $\eta^0 - \eta^G$ in
${\cal L}^{(0)}$).
 
It should be noticed, that there is one dangerous case in the above
general picture. It is when the denominator in (45) is small. Such
situation may take place when $n$ has the meaning of $\eta^{_{Gl}}$,
since only in this case $M_{n}^2 = \mbox{O}(1)$, as well as
$M_{\widetilde\eta^0}^2$ does. Then, due to the `play of numbers' the
difference $M_{\widetilde\eta^0}^2 - M_{\eta^{Gl}}^2$ may take any
value, including one which is close to the value of the numerator in
(45). If this situation does take place, then one has to
reconsider the above analysis, rejecting the approximation scheme
(43)--(47) and making instead numerical estimates.

\section{ Radiative decays }

According to the widely spread opinion, radiative decays are the best
tool for the phenomenological investigation of the PS meson mixing.
Usually, the well-known PCAC formulae are drawn for this purpose.
However, as we pointed out in Introduction, in the singlet channel
the usual PCAC formula was valid no longer. The corrected PCAC
formula was proposed in \cite{S-V}. It involves a new `decay'
constant instead of the usual axial-vector-current one and an
additional proper vertex. (Actually, \cite{S-V} discussed the process
$\eta'\rightarrow\gamma\gamma$ with the interpolating $\eta'$ field
defined on the base of the properly normalized current $\mbox{\sf
J}^0_5 = i\bar q \gamma_5 \lambda^0/2 \,q$. So, the $\eta'$ of
\cite{S-V} and our $\eta^0$ are the same objects. Notice, \cite{S-V}
did not consider, however, the mixings of $\eta'$.)

It would be worth investigating the decay
$\eta^0\rightarrow\gamma\gamma$ in the approach of the chiral
effective lagrangian. In fact, the first result of \cite{S-V}, which
concerns the appearance of a new `decay' constant in the correct
formula, is reproduced trivially. Indeed, when U(1)$_A$ symmetry is
taken into consideration, then the $p^4$-order WZW term, which is
responsible for two-photons decays, must involve the nonet-field
matrix $\Sigma = U\exp (i\lambda^0 \eta^0/F_0)$ instead of the usual
octet matrix $U$. Since $\Sigma$ involves the singlet interpolating
field $\eta^0$ divided by $F_0$, the appearance of a new `decay'
constant is obvious. The second result, which concerns the presence
of an additional proper vertex in the correct formula, may be
reproduced, too. Its origin in our approach is related with the
presence in the chiral effective lagrangian of an additional
chiral-invariant term of order $p^4$, which contains the totally
antisymmetric tensor $\epsilon_{\mu\nu\rho\sigma}$, and which,
notwithstanding, is parity-even,
\begin{equation}
{\cal L}_{eff} \> = \> \dots + 
{\cal L}_{WZW} + 
\upsilon_3 \, \epsilon_{\mu\nu\rho\sigma} \,
< F_L^{\mu\nu}F_L^{\rho\sigma} +
		   F_R^{\mu\nu}F_R^{\rho\sigma} > .
\end{equation}
In (48) dots mean the usual chiral-invariant lagrangian, which is
irrelevant to the discussion, ${\cal L}_{WZW}$ is the WZW lagrangian,
the last term is the very additional one. The quantity $\upsilon_3$
is a chiral-invariant function with positive charge conjugation and
negative parity. So, the parity and the charge conjugation of the
additional term are correct. Moreover, the additional term is RG
invariant as well, since it does not coincide with the external-field
counterterm discussed in \cite{Shore} (see, also, Section 2) and
$\upsilon_3$ depends on RG invariant variables.

It is easy to detect the contribution from the additional term to the
process $\eta^0\rightarrow \gamma\gamma$. Indeed, let the power
expansion of $\upsilon_3$ be
\begin{equation}
\upsilon_3 \> =\> g_0 (\eta^0 + F_0 \Theta) +
                 \sum_{\kappa} g_{\kappa}\eta^{\kappa} + \dots
\end{equation}
with $g_0$ and $g_{\kappa}$ the constants. Substituting (49) into
(48) and extracting only terms with two photons, one may conclude
that $g_0$ is really the proper vertex that contributes to
$\eta^0\rightarrow\gamma\gamma$. However, unlike \cite{S-V}, we have
not any reason to interpret it as the ``coupling of the glue
component of $\eta'$ to photons''. Indeed, $\eta^0$ describes quarkic
degrees of freedom (on the projection to interpolating fields)
whereas gluonic ones are described by $\eta^G$. So, the coupling of
the ``glue'' to photons might rather be peculiar to term
$g_{G}\eta^G$ but not to $g_0\eta^0$. Moreover, in our opinion the
coupling $g_G$ must be equal to zero, since the pure ``glue'' cannot
interact directly with photons.\footnote{ Nevertheless, an indirect
interaction of gluons to photons is allowed. This property is
represented in (49) by the term $g_0 F_0 \Theta$ which describes the
QCD Green function (not the vertex!) of the gluon anomaly operator
and two photons. Notice, the coupling constant of this term coincides
with the vertex $g_0$ due to U(1)$_A$ invariance.} Therefore the term
$g_G\eta^G$ must be suppressed in (49). On the base of this reason we
can think, also, that the ``glue'' proper vertex of \cite{S-V} in the
reality is equal to zero, as well. (Let us emphasize, that the proper
vertex of \cite{S-V} was introduced in rather indirect manner; the
necessity of its presence was not supported by any other observables,
and no arguments were presented why it was nonzero.) Nevertheless,
there is the $\eta^0$-proper vertex. One may detect it in the
framework of the approach of \cite{S-V} if one rejects the
approximation $k^2 = 0$ in equation (4.2) of \cite{S-V} and
instead considers the mass-shell condition $k^2 = M_{\eta'}^2$ with
$M_{\eta'}^2 \not= 0$. Note, although the condition $k^2 = 0$ is
usual in PCAC, it is not an adequate approximation in case of
non-Goldstone states.

It is easy to show that $g_0 F_0 \sim N_c^0$ at large $N_c$.
Therefore $g_0\sim N_c^{-1/2}$. So, taking into account the factor
$N_c^{1/2}$ in the WZW term, one may conclude that $g_0$ contributes
to the amplitude of $\eta^0\rightarrow\gamma\gamma$ in the
next-to-leading order in the large-$N_c$. Nevertheless, the
contribution of the proper vertex is not negligible in the
phenomenological sense. On the contrary, when one considers the real
decay $\eta'\rightarrow\gamma\gamma$, then the proper vertex will
contribute in the same order of the combined chiral and large-$N_c$
expansion in which the mixing $\eta^0 - \eta^8$ contributes. Really,
taking into account the WZW factor $N_c^{1/2}$ and disregarding the
overall factor, the amplitude of $\eta'\rightarrow\gamma\gamma$ may
be represented as
\begin{eqnarray}
A_{\eta' \rightarrow \gamma\gamma} & \propto &
       \Omega^8_{\eta'} + 2\sqrt{2}r\Omega^0_{\eta'} + 
       N_c^{-1/2} g_{0}\Omega^0_{\eta'}                  \\
& = &  s_2 + 2\sqrt{2}\,r c_1 c_2 + N_c^{-1/2} g_{0}c_1 c_2 .\nonumber
\end{eqnarray}
Here we have used (34) and only take into account the contributions
of $\eta^8$ and $\eta^0$ (other ones are really negligible).  Basing
on the results of the previous section one can see that the second
term in the r.h.s is the leading one. It behaves as ${\mbox O}(1)$
and describes the $\eta^0$-contribution that arises from the WZW
term. The first term describes the WZW $\eta^8$-contribution. The
third term is caused by the additional term in (48). Since these
latter two terms behave as O$(p^2N_c)$ and O$(N_c^{-1})$,
respectively, they both belong to one and the same (next-to-leading)
order O$(\varepsilon)$ of the combined chiral and large-$N_c$
expansion.

The above analysis shows that when studding the decay
$\eta'\rightarrow\gamma\gamma$ with the $\eta - \eta'$ mixing is
neglected, one must also neglect the proper vertex $g_{0}$. However,
if the $\eta - \eta'$ mixing is taken into account, then one must
take into account the $\eta^0-\eta^G$ mixing, the proper vertex
$g_0$, and the effect $F_0\not=F$ ($r = 1 + \mbox{O}(N_c^{-1})$),
since all these effects contribute to the amplitude of the decay in
one and the same order of the combined chiral and large-$N_c$
expansion. On the contrary, the PCAC formula for
$\eta\rightarrow\gamma\gamma$ works well with the $\eta - \eta'$
mixing taken into account, since the additional term in (48) and the
effect $F_0\not=F$ contribute to the amplitude of the decay
$\eta\rightarrow\gamma\gamma$ starting at order O$(\varepsilon^{2})$,
i.e. in the next-next-to-leading order whereas the contribution of
the $\eta^8 - \eta^0$ mixing is of order O$(\varepsilon)$.

\section{ Summary and discussion }

The present paper proposes a model-independent way to constrain the
contributions of the singlet PS states to the chiral effective
lagrangian with an accounting their nature. This allows one to
independently identify the singlet states in the framework of the
approach, and it opens the way to systematic investigation of the
low-laying singlet resonances in QCD. Moreover, it becomes possible
to interpret more precisely the results of some previous
investigations. For instance, one should conclude that the PS
`glueball' of \cite{Gounaris} is not really a glueball but rather an
excitation of the singlet quarkic state, because in \cite{Gounaris}
it was included into lagrangian ${\cal L}^{(mass)}$ which involved
the current quark masses and, therefore, explicitly broke the flavor
symmetry.

A special attention is paid in the present paper to construct a
correct generalization of the chiral effective lagrangian which would
involve singlet interpolating fields and satisfy not only the chiral
symmetry but the QCD-inspired RG symmetry, too. Correct account of RG
symmetry allows us to introduce the gluonic and singlet quarkic
interpolating fields to be RG invariant objects which separately
describe the gluonic and singlet quarkic degrees of freedom in the
effective theory. Owing to dynamical reasons these interpolating
fields may mix in the lagrangian. However, this mixing remains RG
invariant in spite of the fact that the relevant composite operators
in QCD mix under RG.  This property shows a certain advantage in
describing singlet states in the framework of the effective theory.

Besides the above results, which have rather general significance,
the present paper also proposes some particular results. Thus the
mixing among the iso-singlet PS states is investigated and the
hierarchy of the mixing angles is obtained which is defined by the
combined chiral and large-$N_c$ expansion. The largest mixing angle
is seen to be between the singlet quarkic lowest state $\eta^0$ and
the gluonic ground-state $\eta^G$. Then, in order of decreasing
significance, follow the $\eta^8 - \eta^0$ mixing, and the $\eta^0 -
\widetilde\eta^0$, $\eta^8 - \widetilde\eta^0$ mixings, where
$\widetilde\eta^0$ is the excitation of $\eta^0$. The mixing angles
$\eta^G - \eta^8$, $\eta^G - \widetilde\eta^0$ turn out to be the
smallest ones and negligible in the approximation up to and including
O($\varepsilon^2$) where $\varepsilon$ is the parameter of the
expansion.

Another important application concerns the radiative decays
$\eta\rightarrow\gamma\gamma$ and $\eta'\rightarrow\gamma\gamma$. We
reproduce the modern PCAC results \cite{S-V}, i.e. we show that the
correct formula for $\eta'\rightarrow\gamma\gamma$ must involve a
special `decay' constant instead of the usual axial-vector-current
decay constant, and an additional proper vertex. However, the nature
of the proper vertex is found different in our approach as compared
with that of \cite{S-V}. Besides, we show that the proper vertex
contributes to the amplitude of the decay
$\eta'\rightarrow\gamma\gamma$ in the same order of the combined
chiral and large-$N_c$ expansion in which the $\eta^8$-state
contributes due to the $\eta - \eta'$ mixing. Therefore, the effect
of the proper vertex must be considered together with the mixing. On
the contrary, the well-known PCAC formula for the decay
$\eta\rightarrow\gamma\gamma$ works well without any modifications,
even when the mixing $\eta - \eta'$ is taken into consideration. So,
this decay remains a good tool for the study of the mixing.
(Nevertheless, the proper vertex contributes to the amplitude of
$\eta\rightarrow\gamma\gamma$ in the next-to-leading order where the
octet decay constant $F$ is split, $F_8 \not= F_{\pi}$. So, the
$p^4$-order corrections to $\eta\rightarrow\gamma\gamma$ must be
taken into account together with the proper vertex.)

On the whole, the results of this paper establish a formal framework
which to perform quantitative estimates. Such work should take into
account the real spectrum of the observed mesons and the data on
their decays. One might obtain the quantitative description of the
mixing as the output result, which is necessary for interpretation of
the nature of the observed mesons. Of course, an extension of the
analysis to other problems and to other channels would be possible.
These questions will be addressed in forthcoming papers.

\begin{flushleft}
{\large \bf Acknowledgements}
\end{flushleft}

The author is grateful to B.A.Arbuzov for discussions. The work was
supported in part by grant RFBR 95-02-03704-$a$.

\end{document}